\documentclass[conference]{IEEEtran}
\IEEEoverridecommandlockouts

\usepackage{cite}
\usepackage{amsmath,amssymb,amsfonts}
\usepackage{graphicx}
\usepackage{textcomp}
\usepackage{xcolor}
\def\BibTeX{{\rm B\kern-.05em{\sc i\kern-.025em b}\kern-.08em
    T\kern-.1667em\lower.7ex\hbox{E}\kern-.125emX}}

\usepackage{tikz,pgfplots}
\usepgfplotslibrary{groupplots,units}
\usetikzlibrary{patterns}
\pgfplotsset{compat=newest}
\usetikzlibrary{shapes.geometric, arrows, calc, fit, shapes, backgrounds}
\usepgfplotslibrary{colorbrewer}

\usepackage{pifont}
\newcommand{\xmark}{\ding{55}}

\usepackage{acronym}
\usepackage{csquotes}
\usepackage{hyperref}
\usepackage{balance}
\usepackage[capitalize,nameinlink]{cleveref}

\crefname{figure}{Fig.\@}{Figs.\@}
\Crefname{figure}{Figure\@}{Figures\@}

\crefname{table}{Table\@}{Tables\@}
\Crefname{table}{Table\@}{Tables\@}

\usepackage{adjustbox}
\usepackage{nicematrix}
\usepackage{booktabs}
\usepackage{multirow}
\usepackage{threeparttable}

\acrodef{act}[ACT]{auxiliary classification task}
\acrodef{asd}[ASD]{anomalous sound detection}
\acrodef{knn}[k-NN]{k-nearest neighbors}
\acrodef{ldn}[LDN]{local density-based anomaly score normalization}
\acrodef{ced}[CED]{cluster exit detection}
\acrodef{varmin}[VarMin]{variance minimization}

\begin{document}

\title{How Much Does Machine Identity Matter in Anomalous Sound Detection at Test Time?
}

\author{\IEEEauthorblockN{\textit{Kevin Wilkinghoff$\,^{1,2}$, Keisuke Imoto$^{3}$, Zheng-Hua Tan$^{1,2}$}\vspace{.7\baselineskip}}
\IEEEauthorblockA{$^{1}~$Aalborg University, Denmark, $^{2}~$Pioneer Centre for Artificial Intelligence, Denmark, $^{3}~$Kyoto University, Japan}
}

\maketitle

\begin{abstract}
\Ac{asd} benchmarks typically assume that the identity of the monitored machine is known at test time and that recordings are evaluated in a machine-wise manner.
However, in realistic monitoring scenarios with multiple known machines operating concurrently, test recordings may not be reliably attributable to a specific machine, and requiring machine identity imposes deployment constraints such as dedicated sensors per machine. To reveal performance degradations and method-specific differences in robustness that are hidden under standard machine-wise evaluation, we consider a minimal modification of the \ac{asd} evaluation protocol in which test recordings from multiple machines are merged and evaluated jointly without access to machine identity at inference time. Training data and evaluation metrics remain unchanged, and machine identity labels are used only for post hoc evaluation. Experiments with representative \ac{asd} methods show that relaxing this assumption reveals performance degradations and method-specific differences in robustness that are hidden under standard machine-wise evaluation, and that these degradations are strongly related to implicit machine identification accuracy.
\end{abstract}


\acresetall

\section{Introduction}
\label{sec:intro}

\Ac{asd} aims to detect unusual sounds and is widely used in machine condition monitoring.
The DCASE Challenge \cite{mesaros2025decade} has become a central benchmark for \ac{asd}.
Recent editions emphasize robustness to domain shifts \cite{kawaguchi2021description,dohi2022description,dohi2023description,nishida2024description,nishida2025description}, motivating extensive work on domain adaptation and generalization \cite{wilkinghoff2025handling}.
Despite these advances, current \ac{asd} evaluations rely on an implicit assumption: Test recordings are associated with a known machine identity and evaluated in a machine-wise manner.
While the \emph{first-shot} setting \cite{dohi2023description,nishida2024description,nishida2025description} reduces overly optimistic evaluation by enforcing disjoint machine types between system development and evaluation, it still assumes access to machine identity at test time.

This assumption does not always hold in practice. In realistic monitoring scenarios, multiple known machines may operate concurrently, and test recordings cannot always be reliably attributed to a specific machine. From an operator perspective, it is therefore desirable for a system to identify which machine behaves abnormally, rather than assuming that machine identity is given. One alternative is to use microphone arrays and sound source localization to infer the originating machine, but such solutions increase cost, system complexity, and installation effort, and are often impractical in existing industrial settings. Moreover, relying on dedicated sensing infrastructure for each machine limits flexibility when machine layouts or inventories change. Evaluating \ac{asd} without machine identity therefore supports more scalable and reusable monitoring systems that can be deployed across different factories and operating conditions.

These practical constraints are not reflected in current evaluation protocols. Many approaches rely on machine-specific models \cite{harada2023first-shot}, machine-specific operating-condition information \cite{wilkinghoff2021combining,chen2022self-supervised}, or machine-specific test-set statistics for anomaly score normalization \cite{saengthong2025deep}. All of these require access to machine identity or delayed aggregation of test data. Such assumptions may not hold in realistic monitoring scenarios, causing standard evaluations to hide robustness differences that only emerge when they are violated.

Throughout this work, we consider the same scope assumptions as current DCASE benchmarks: single-channel recordings that contain sounds from a single machine, and a fixed, known set of machines. The only assumption that is relaxed is the availability of machine identity at test time.
Test recordings from multiple known machines are merged into a single test set and evaluated jointly, while training data and evaluation metrics remain unchanged.
Machine identity labels are used only for post hoc evaluation.
Rather than proposing new detection models, our goal is to analyze how representative \ac{asd} methods behave when this single evaluation assumption is relaxed.
The contributions of this work are:
\begin{itemize}
    \item We make explicit the implicit assumption of known machine identity in current \ac{asd} evaluation protocols;
    \item We propose a minimal evaluation protocol that removes machine identity at inference time while keeping all other aspects unchanged;
    \item We empirically show that relaxing this assumption exposes method-specific robustness differences and link anomaly detection performance degradation to implicit machine identification accuracy, quantified via an auxiliary post hoc identification task.
\end{itemize}

\section{Background and Problem Setting}
\label{sec:background}

We briefly summarize the standard \ac{asd} formulation and evaluation protocol in the DCASE Challenge to make explicit the assumptions relevant to this work and position our setting relative to existing benchmarks.

\subsection{Anomalous Sound Detection Setup}
\label{sec:asd_setup}

In acoustic machine condition monitoring, \ac{asd} aims to detect sounds that deviate from a machine’s normal operating behavior, typically under the assumption that anomalous examples are rare or unavailable during training. The DCASE Challenge on unsupervised \ac{asd} \cite{koizumi2020description,kawaguchi2021description,dohi2022description,dohi2023description,nishida2024description,nishida2025description} has become the de facto benchmark for this task.

In the standard DCASE setting, datasets are organized into development and evaluation splits, each containing machine-specific training and test data. Training data consist exclusively of normal recordings from individual machines and are associated with machine IDs. At evaluation time, test recordings are grouped by machine, include both normal and anomalous samples, and performance is reported using machine-wise metrics. Models are therefore trained either separately for each machine or using a shared model with access to machine identity at inference time.

\subsection{Implicit Assumption of Known Machine Identity}
\label{sec:machine_id_assumption}

An implicit but central assumption in this protocol is that machine identity is known at test time. Each test recording is assumed to be associated with a specific machine ID, and anomaly scores are computed and evaluated within machine-wise partitions. While this assumption simplifies both modeling and evaluation, it does not necessarily hold in practical monitoring scenarios, where multiple known machines may operate concurrently and recordings cannot always be reliably attributed to a specific machine. In the remainder of this paper, we analyze the consequences of relaxing this assumption.

\section{Evaluation Protocol Without Machine Identity}
\label{sec:eval_no_id}

\begin{figure}
    \centering
    \begin{adjustbox}{width=\columnwidth}
          \includegraphics{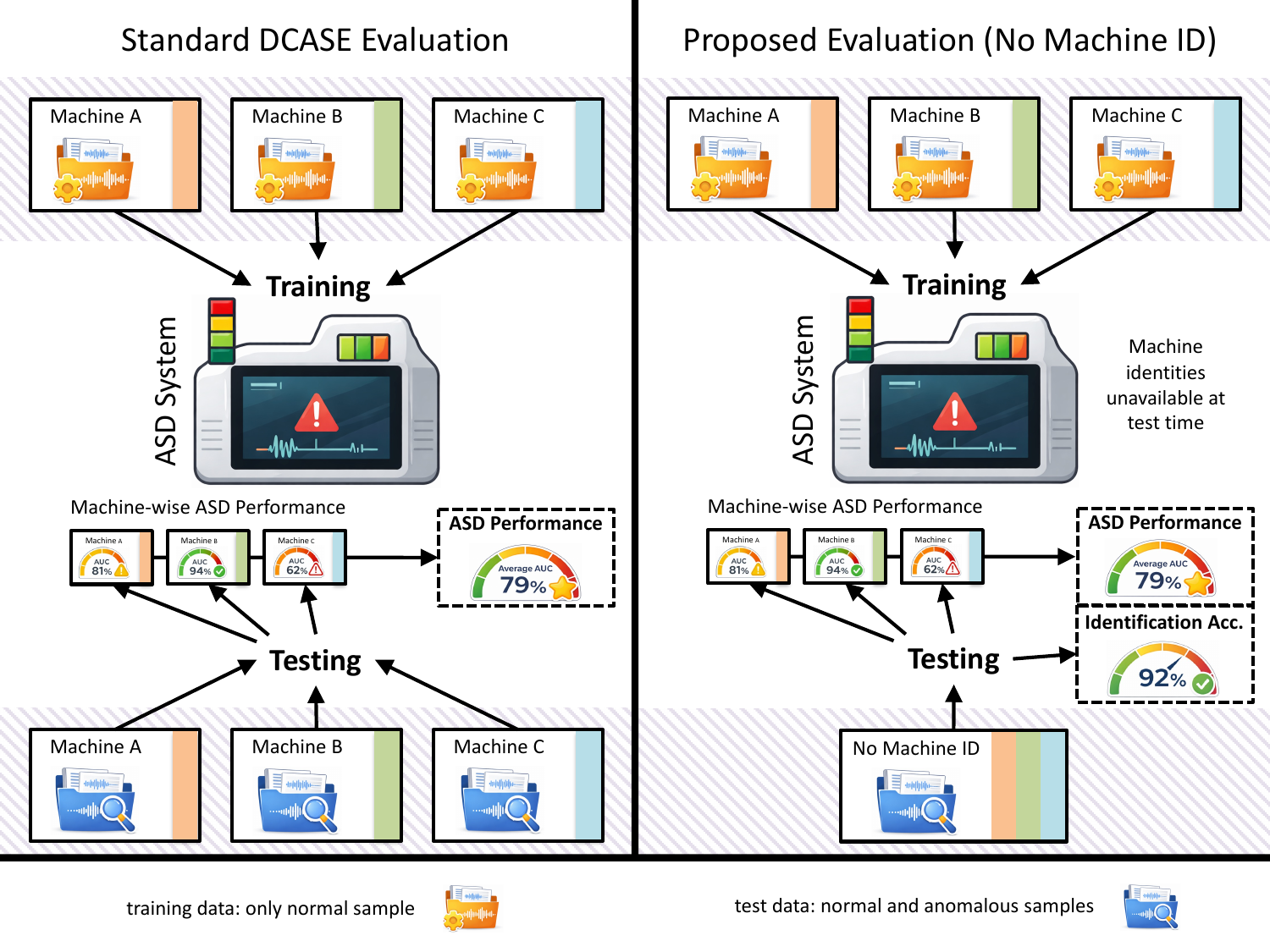}
    \end{adjustbox}
    \vspace*{-0.8cm}
    \caption{Comparison of the standard DCASE evaluation protocol and the proposed evaluation without machine identity. Training data and anomaly detection metrics are unchanged; machine identity is unavailable at inference time, and machine identification accuracy is reported as an auxiliary metric.}
    \label{fig:setup}
\end{figure}

We introduce a minimal modification of the standard DCASE evaluation protocol in which machine identity is unavailable at test time, while all other aspects remain unchanged. This isolates the impact of machine identity on \ac{asd} performance. \Cref{fig:setup} compares the standard setting with the proposed identity-free variant.

\subsection{Test Set Construction}
\label{sec:test_set_construction}

For both the development and evaluation splits of the dataset (cf. \cref{sec:asd_setup}), all test recordings from multiple known machines are merged into a single test set, spanning different machine types. The original development–evaluation split and label distributions are preserved. Machine identity information is removed at inference time, while the audio recordings themselves are used as-is.

\subsection{Tasks and Outputs}
\label{sec:tasks_outputs}

The primary task remains anomaly detection, defined exactly as in the standard \ac{asd} setting: Given an input audio recording, the system produces an anomaly score indicating the likelihood that the recording deviates from normal operation. Models are not informed of the machine identity and must process all test recordings jointly.

We additionally consider a secondary task in which a method predicts the machine ID of a test recording. This task is not part of the main \ac{asd} objective and is evaluated only after inference using machine identity labels that are unavailable to the model at test time. Beyond serving as a diagnostic tool, it is also practically motivated by monitoring scenarios in which identifying the abnormal machine is required.

We now formalize the relationship between anomaly detection without machine identity and implicit machine identification. Let $s_m(x)$ denote a machine-specific anomaly score for a test recording $x$, obtained either from a machine-specific model or implicitly via reference data. When machine identity is known, detection is based on $s_{m^\ast}(x)$, where $m^\ast$ is the true machine ID. In the absence of machine identity, a natural aggregation strategy is is minimum aggregation,
\begin{equation}
s(x) = \min_{m} s_m(x),
\end{equation}
corresponding to selecting the machine whose notion of normality best explains the test recording \cite{tax2004support}.
In contrast to open-set classification, the machine set is fixed and known, and the challenge lies in resolving the correct machine-specific normality model without identity labels, rather than handling unseen classes.
Without access to machine identity at inference time, anomaly detection performance matches the standard protocol precisely when the implicitly selected machine ID
\begin{equation}
\arg\min_{m} s_m(x) = m^\ast \quad \text{for all } x,
\end{equation}
highlighting that anomaly detection without machine identity implicitly requires correct machine identification. Consequently, any degradation in anomaly detection performance relative to the standard setting arises from incorrect implicit machine identification and is expected to increase as the number of candidate machines grows. For rank-based metrics such as the area under the ROC curve (AUC) and partial AUC (pAUC) \cite{mcclish1989analyzing,fawcett2006introduction}, this degradation occurs only when an incorrect machine is selected and thus grows with the probability of such errors,
\begin{equation}
\mathbb{P}\!\left(\arg\min_{m} s_m(x) \neq m^\ast\right).
\end{equation}

\subsection{Evaluation Metrics}
\label{sec:metrics}

In contrast to the modified test set construction described above, evaluation itself follows the standard DCASE protocol. After inference, anomaly scores are retrospectively associated with machine IDs and evaluated per machine using AUC and pAUC with $p=0.1$, followed by averaging across machines as in the original benchmark.

In addition to anomaly detection performance, we report implicit machine identification accuracy, which is chance-normalized to account for differing numbers of machines across datasets and defined as $(a - 1/K)/(1 - 1/K)$, where $a$ denotes the raw identification accuracy and $K$ the number of machines. 
To relate anomaly detection performance degradation to implicit machine identification accuracy, we define a chance-normalized degradation of anomaly detection performance. Since both AUC and pAUC have a fixed chance level of $0.5$, the normalized degradation is given by
\begin{equation}
\Delta_{\text{norm}}
=
1 - \frac{ A_{\text{unknown}} - 0.5}{A_{\text{known}} - 0.5},
\end{equation}
where $A_{\text{known}}$ and $A_{\text{unknown}}$ denote anomaly detection performance with known and unknown machine identity, respectively, computed using the standard DCASE evaluation protocol.
The measure is undefined when $A_{\text{known}} \leq 0.5$, i.e., when no discriminative performance above chance is available.
This quantity represents the fraction of performance above chance lost when machine identity is unavailable and is directly comparable to chance-normalized machine identification accuracy.
Machine identity labels are used for evaluation and are never provided to the models during inference. Any performance differences relative to the standard protocol can therefore be attributed solely to the absence of machine identity information at test time.
Throughout the paper, \enquote{identification accuracy} denotes implicit machine identification via score aggregation, not an explicit classifier.

\section{Experimental Setup}

In the previous section, we introduced an evaluation protocol without machine identity at test time. Here, we quantify the resulting performance changes across different \ac{asd} systems.

\subsection{Datasets}
We conduct our experimental evaluation on five publicly available datasets for semi-supervised acoustic anomaly detection from the DCASE Challenge. Specifically, we consider DCASE2020 \cite{koizumi2020description}, constructed from MIMII \cite{purohit2019mimii} and ToyADMOS \cite{koizumi2019toyadmos}; DCASE2022 \cite{dohi2022description}, based on MIMII-DG \cite{dohi2022mimiidg} and ToyADMOS2 \cite{harada2021toyadmos2}; DCASE2023 \cite{dohi2023description}, extending MIMII-DG and ToyADMOS2+ \cite{harada2023toyadmos2+}; DCASE2024 \cite{nishida2024description}, incorporating MIMII-DG, ToyADMOS2\# \cite{niizumi2024toyadmos2sharp}, and recordings collected under the IMAD-DS setup \cite{albertini2024imadds}; and DCASE2025 \cite{nishida2025description}, consisting of MIMII-DG, ToyADMOS2025 \cite{harada2025toyadmos2025}, and IMAD-DS.

Across these datasets, each development and evaluation set contains between 7 and 21 machines belonging to 6–7 machine types. Except for DCASE2020, which contains a single domain, all datasets provide $990$ source-domain and $10$ target-domain training samples per machine. In the test sets, samples from different domains are balanced, but explicit domain labels are not provided.
For both the development and evaluation splits, test recordings are merged per split following the procedure described in \cref{sec:test_set_construction}. Results are computed separately for development and evaluation splits, and reported individually (\cref{fig:scatter}) and aggregated across splits and datasets following the standard DCASE averaging protocol (\cref{tab:system_res}).

\subsection{\acs{asd} Systems}

We evaluate a diverse set of \ac{asd} systems spanning three commonly used modeling paradigms \cite{wilkinghoff2024audio}:
(i) discriminatively trained models that learn to distinguish between machine identities or operating states during training,
(ii) training-free approaches based on pre-trained embeddings, including OpenL3 \cite{cramer2019look}, BEATs \cite{chen2023beats}, EAT \cite{chen2024eat}, and Dasheng \cite{dinkel2024dasheng}, and
(iii) methods relying on machine-specific models or reference data.
These paradigms are not intended to be exhaustive, but to cover representative and widely used classes of \ac{asd} approaches in the DCASE literature.
In particular, many generative-based methods ultimately produce machine-specific anomaly scores and fall into category (iii) under the proposed evaluation setting.
Owing to space constraints, we omit descriptions of individual methods. All systems are evaluated exactly as originally proposed, reusing the reported feature extraction pipelines and hyperparameters, without architectural modifications or tuning for the proposed evaluation setting.

In addition, we assess the effect of commonly used anomaly score normalization strategies that are designed to improve robustness under domain shifts. Specifically, we consider normalization approaches that adjust anomaly scores using machine-specific reference samples prior to score computation \cite{wilkinghoff2025keeping,wilkinghoff2025local,matsumoto2025adjusting}. These normalization methods are applied on top of the evaluated \ac{asd} systems and analyzed with respect to both anomaly detection performance and implicit machine identification accuracy.

\section{Results}
We analyze robustness across \ac{asd} systems and relate detection degradation to implicit machine identification accuracy.

\subsection{Robustness of ASD Systems to Unknown Machine Identity}
\label{sec:system_res}
\begin{table*}[!t]
\centering
\caption{Average anomaly detection performance with and without access to machine identity, together with chance-normalized implicit machine identification accuracy, for different \ac{asd} systems across the development and evaluation sets of the DCASE2020, DCASE2022, DCASE2023, DCASE2024, and DCASE2025 datasets.}
\begin{adjustbox}{max width=\textwidth,max totalheight=\textheight}
\begin{NiceTabular}{l *{2}{c} *{4}{c}}
\toprule
&\multicolumn{2}{c}{\textbf{Score normalization}}& \multicolumn{3}{c}{\textbf{\ac{asd} performance}} & \\
\cmidrule(lr){2-3}\cmidrule(lr){4-6}
\textbf{\acs{asd} system}&\acs{ldn} \cite{wilkinghoff2025keeping,wilkinghoff2025local}&\acs{varmin} \cite{matsumoto2025adjusting}& known ID & unknown ID & $\Delta_\text{norm}$ & \textbf{ID acc. (norm.)} \\
\midrule

\multicolumn{5}{l}{\textit{Discriminative models}} \\
Direct-ACT \cite{wilkinghoff2025keeping,wilkinghoff2025local} & \xmark & & $70.31\%$ & $69.66\%$ & $3.20\%$ & $88.85\%$ \\
Direct-ACT \cite{wilkinghoff2025keeping,wilkinghoff2025local} (variant) & \xmark & \xmark & $70.83\%$ & $70.35\%$ & $2.30\%$ & $89.31\%$ \\
Direct-ACT \cite{wilkinghoff2025keeping,wilkinghoff2025local} (variant) &  &  & $66.02\%$ & $65.76\%$ & $1.62\%$ & $92.59\%$ \\

\midrule
\multicolumn{5}{l}{\textit{Training-free embedding-based approaches}} \\
openL3-based \cite{wilkinghoff2026temporal}& \xmark & \xmark & $65.44\%$ & $63.51\%$ & $12.50\%$ & $81.61\%$ \\
BEATs-based \cite{wilkinghoff2026temporal} & \xmark & \xmark & $68.71\%$ & $67.30\%$ & $7.54\%$ & $84.23\%$ \\
BEATs-based \cite{wilkinghoff2026temporal} (variant) & \xmark &  & $67.29\%$ & $65.38\%$ & $11.05\%$ & $80.55\%$ \\
BEATs-based \cite{wilkinghoff2026temporal} (variant) &  &  & $64.47\%$ & $63.80\%$ & $4.63\%$ & $90.53\%$ \\
EAT-based \cite{wilkinghoff2026temporal} & \xmark & \xmark & $65.69\%$ & $63.53\%$  & $13.77\%$ & $74.20\%$ \\
Dasheng-based \cite{wilkinghoff2026temporal} & \xmark & \xmark & $64.59\%$ & $62.98\%$ & $11.03\%$ & $82.43\%$ \\

\midrule
\multicolumn{5}{l}{\textit{Machine-specific models}} \\
autoencoder \cite{koizumi2020description} (re-implementation)&& & $56.71\%$ & $55.40\%$ & $19.52\%$ & $86.43\%$ \\

\bottomrule
\end{NiceTabular}

\end{adjustbox}
\label{tab:system_res}
\end{table*}

\Cref{tab:system_res} summarizes average anomaly detection performance, measured by AUC and pAUC and averaged as in the DCASE protocol, with and without access to machine identity across all evaluated datasets. In addition to absolute performance, the table reports chance-normalized anomaly detection degradation and chance-normalized implicit machine identification accuracy. Discriminatively trained models exhibit the smallest relative performance degradation, indicating strong robustness when machine identity is unavailable. In contrast, training-free methods based on pre-trained embeddings or machine-specific models show substantially larger drops, reflecting their sensitivity to incorrect implicit machine selection.

A noteworthy exception is the behavior of \ac{ldn}. Although it substantially reduces implicit machine identification accuracy, it simultaneously improves absolute anomaly detection performance.
In fact, even without access to machine identity at test time, systems using \ac{ldn} outperform their counterparts evaluated with known machine identity but without \ac{ldn}. This observation indicates that high identification accuracy is not a strict prerequisite for strong anomaly detection performance. Robustness to unknown machine identity also depends on how anomaly scores are structured and aggregated, not solely on how reliably machines can be identified.

\subsection{Relation between Identification and Detection Performance}
\begin{figure}[!t]
\centering
\begin{tikzpicture}
\begin{axis}[
    width=\columnwidth,
    height=0.63\columnwidth,
    xlabel={Chance-normalized identification accuracy (\%)},
    ylabel={Chance-normalized \acs{asd} degradation (\%)},
    xmin=49, xmax=100,
    ymin=-1, ymax=50,
    axis y line*=left,
    axis x line*=bottom,
    grid=major,
    major grid style={line width=.2pt, draw=gray!40},
    tick align=outside,
    tick style={black},
    cycle list/Paired,
    legend style={
        at={(0.5,1.08)},
        anchor=south,
        fill=none,
        font=\scriptsize,
        legend image post style={scale=0.8},
        legend columns=2,
        column sep=4pt,
    }
]

\addplot+[
    only marks,
    mark=*,
    mark size=2.6pt,
]
coordinates {
(78.54, {100*(1-(91.4-50)/(92.77-50))})  
(67.79, {100*(1-(91.50-50)/(92.43-50))})  
(82.22, {100*(1-(89.80-50)/(91.02-50))})  
(66.45, {100*(1-(75.68-50)/(80.54-50))})  
(70.58, {100*(1-(82.75-50)/(86.84-50))})  
(64.99, {100*(1-(78.84-50)/(84.45-50))})  
(78.00, {100*(1-(72.94-50)/(75.43-50))})  
(66.12, {100*(1-(75.35-50)/(79.10-50))})  
(66.85, {100*(1-(75.63-50)/(80.87-50))})  
(70.39, {100*(1-(62.79-50)/(67.69-50))})  
};
\addlegendentry{DCASE2020 (dev)}

\addplot+[
    only marks,
    mark=*,
    mark size=2.6pt,
]
coordinates {
(68.67, {100*(1-(90.90-50)/(91.36-50))})  
(71.84, {100*(1-(91.54-50)/(92.03-50))})  
(79.38, {100*(1-(90.31-50)/(90.65-50))})  
(65.27, {100*(1-(75.39-50)/(81.76-50))})  
(68.29, {100*(1-(83.69-50)/(87.39-50))})  
(63.67, {100*(1-(80.42-50)/(85.32-50))})  
(74.78, {100*(1-(75.43-50)/(77.83-50))})  
(56.34, {100*(1-(74.26-50)/(82.21-50))})  
(65.51, {100*(1-(75.84-50)/(81.60-50))})  
(68.41, {100*(1-(65.55-50)/(71.75-50))})  
};
\addlegendentry{DCASE2020 (eval)}

\addplot+[
    only marks,
    mark=square*,
    mark size=2.6pt,
]
coordinates {
(75.65, {100*(1-(70.95-50)/(73.29-50))})  
(78.60, {100*(1-(72.35-50)/(75.15-50))})  
(83.88, {100*(1-(71.95-50)/(71.94-50))})  
(54.90, {100*(1-(57.95-50)/(61.87-50))})  
(68.35, {100*(1-(65.11-50)/(66.14-50))})  
(60.70, {100*(1-(63.62-50)/(65.08-50))})  
(80.88, {100*(1-(64.60-50)/(64.93-50))})  
(60.57, {100*(1-(59.79-50)/(63.18-50))})  
(60.58, {100*(1-(59.18-50)/(60.54-50))})  
(72.28, {100*(1-(50.89-50)/(51.28-50))})  
};
\addlegendentry{DCASE2022 (dev)}

\addplot+[
    only marks,
    mark=square*,
    mark size=2.6pt,
]
coordinates {
(71.30, {100*(1-(66.15-50)/(68.49-50))})  
(78.83, {100*(1-(69.09-50)/(69.68-50))})  
(82.08, {100*(1-(63.89-50)/(64.93-50))})  
(52.60, {100*(1-(60.67-50)/(64.04-50))})  
(62.88, {100*(1-(62.28-50)/(65.09-50))})  
(57.45, {100*(1-(60.61-50)/(63.78-50))})  
(72.60, {100*(1-(60.57-50)/(62.01-50))})  
(51.02, {100*(1-(60.42-50)/(62.50-50))})  
(55.12, {100*(1-(60.10-50)/(61.81-50))})  
(62.37, {100*(1-(52.54-50)/(53.98-50))})  
};
\addlegendentry{DCASE2022 (eval)}

\addplot+[
    only marks,
    mark=triangle*,
    mark size=2.8pt,
]
coordinates {
(95.67, {100*(1-(71.49-50)/(71.53-50))})  
(97.08, {100*(1-(71.79-50)/(71.82-50))})  
(98.50, {100*(1-(65.07-50)/(65.15-50))})  
(86.00, {100*(1-(61.70-50)/(62.20-50))})  
(86.42, {100*(1-(63.65-50)/(65.68-50))})  
(70.75, {100*(1-(60.42-50)/(64.37-50))})  
(99.17, {100*(1-(63.40-50)/(63.48-50))})  
(85.25, {100*(1-(62.20-50)/(62.94-50))})  
(87.83, {100*(1-(60.88-50)/(62.39-50))})  
(91.92, {100*(1-(54.43-50)/(54.64-50))})  
};
\addlegendentry{DCASE2023 (dev)}

\addplot+[
    only marks,
    mark=triangle*,
    mark size=2.8pt,
]
coordinates {
(99.42, {100*(1-(71.52-50)/(71.52-50))})  
(99.50, {100*(1-(71.10-50)/(71.10-50))})  
(100, {100*(1-(57.94-50)/(57.94-50))})  
(96.58, {100*(1-(65.14-50)/(65.20-50))})  
(99.25, {100*(1-(70.90-50)/(70.84-50))})  
(97.00, {100*(1-(69.22-50)/(69.24-50))})  
(99.83, {100*(1-(64.60-50)/(64.55-50))})  
(87.75, {100*(1-(64.71-50)/(65.63-50))})  
(97.92, {100*(1-(63.92-50)/(63.96-50))})  
(99.67, {100*(1-(57.08-50)/(57.08-50))})  
};
\addlegendentry{DCASE2023 (eval)}

\addplot+[
    only marks,
    mark=diamond*,
    mark size=2.8pt,
]
coordinates {
(99.83, {100*(1-(64.47-50)/(64.48-50))})  
(99.92, {100*(1-(64.63-50)/(64.63-50))})  
(99.92, {100*(1-(59.44-50)/(59.45-50))})  
(97.67, {100*(1-(58.06-50)/(58.05-50))})  
(96.17, {100*(1-(58.75-50)/(58.71-50))})  
(96.00, {100*(1-(57.42-50)/(57.35-50))})  
(100, {100*(1-(58.42-50)/(58.42-50))})  
(77.25, {100*(1-(57.51-50)/(58.90-50))})  
(93.58, {100*(1-(56.65-50)/(57.12-50))})  
(99.5, {100*(1-(54.64-50)/(54.67-50))})  
};
\addlegendentry{DCASE2024 (dev)}

\addplot+[
    only marks,
    mark=diamond*,
    mark size=2.8pt,
]
coordinates {
(100, {100*(1-(55.07-50)/(55.07-50))})  
(100, {100*(1-(56.82-50)/(56.82-50))})  
(100, {100*(1-(53.64-50)/(53.64-50))})  
(100.0, {100*(1-(61.01-50)/(61.01-50))})  
(100.0, {100*(1-(60.09-50)/(60.09-50))})  
(99.37, {100*(1-(59.91-50)/(59.94-50))})  
(100, {100*(1-(55.93-50)/(55.93-50))})  
(76.25, {100*(1-(60.74-50)/(60.92-50))})  
(100, {100*(1-(58.51-50)/(58.51-50))})  
(100, {100*(1-(51.92-50)/(51.92-50))})  
};
\addlegendentry{DCASE2024 (eval)}

\addplot+[
    only marks,
    mark=pentagon*,
    mark size=2.8pt,
]
coordinates {
(99.75, {100*(1-(57.66-50)/(57.66-50))})  
(99.58, {100*(1-(59.96-50)/(59.95-50))})  
(99.92, {100*(1-(56.36-50)/(56.30-50))})  
(96.67, {100*(1-(60.65-50)/(60.81-50))})  
(93.50, {100*(1-(63.47-50)/(63.95-50))})  
(96.67, {100*(1-(60.92-50)/(61.02-50))})  
(100, {100*(1-(63.15-50)/(63.15-50))})  
(84.92, {100*(1-(61.91-50)/(63.04-50))})  
(97.00, {100*(1-(61.21-50)/(61.25-50))})  
};
(99.83, {100*(1-(53.46-50)/(53.46-50))})  
\addlegendentry{DCASE2025 (dev)}

\addplot+[
    only marks,
    mark=pentagon*,
    mark size=2.8pt,
]
coordinates {
(99.71, {100*(1-(56.95-50)/(56.95-50))})  
(100, {100*(1-(54.71-50)/(54.71-50))})  
(100, {100*(1-(49.15-50)/(49.15-50))})  
(100.0, {100*(1-(58.87-50)/(58.87-50))})  
(96.86, {100*(1-(62.29-50)/(62.39-50))})  
(98.86, {100*(1-(62.38-50)/(62.38-50))})  
(100, {100*(1-(58.94-50)/(58.94-50))})  
(96.57, {100*(1-(58.43-50)/(58.47-50))})  
(99.93, {100*(1-(57.87-50)/(57.87-50))})  
(99.93, {100*(1-(50.65-50)/(50.66-50))})  
};
\addlegendentry{DCASE2025 (eval)}

\addplot[
    domain=0:100,
    samples=2,
    dashed,
    thick,
    black
]
{-0.483*x + 47.76};

\end{axis}
\end{tikzpicture}
\caption{Relation between chance-normalized machine identification accuracy and anomaly detection performance degradation across \acs{asd} systems. The dashed line shows a global least-squares fit over all plotted data points. Dataset-specific fits (not shown) exhibit negative slopes, with reduced magnitude when identification accuracy is already near its ceiling for most systems.}
\label{fig:scatter}
\end{figure}
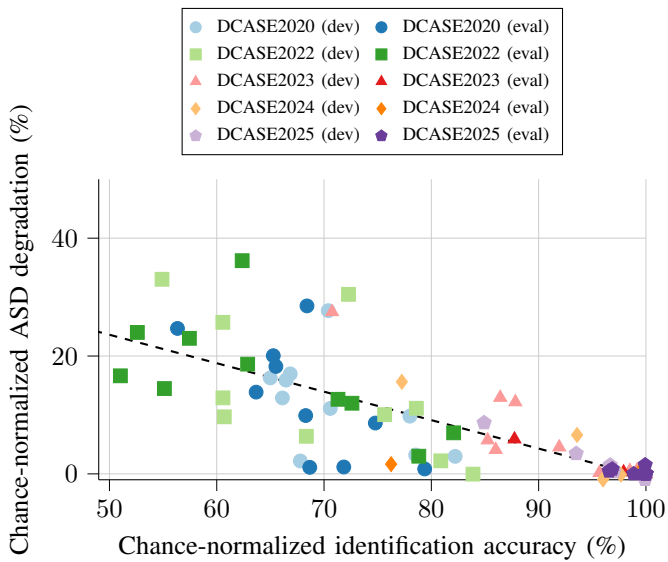

To directly test the relationship suggested by the analysis in \Cref{sec:system_res}, we jointly examine anomaly detection performance degradation and implicit machine identification accuracy across all evaluated \ac{asd} systems. The results, shown in \Cref{fig:scatter}, support the formalization in \Cref{sec:tasks_outputs}, which predicts increasing detection degradation as the probability of selecting an incorrect machine increases.

\section{Discussion}

Removing access to machine identity at test time reveals robustness differences that are largely hidden under standard machine-wise evaluation. While all \ac{asd} systems necessarily rely on machine-specific reference data to define normal behavior, the key distinction is not whether such information is used, but how reliably a method can recover the correct machine-specific notion of normality when machine identity is unavailable. In this regard, discriminatively trained models exhibit particularly strong robustness. Because they are explicitly trained to discriminate between machine identities (and, where applicable, operating conditions), they are less sensitive to background noise and other nuisance factors that do not convey machine identity \cite{wilkinghoff2024why}, resulting in smaller performance degradations when machine labels are removed. In contrast, methods relying on machine-specific models or reference statistics are more vulnerable, as incorrect implicit machine selection directly affects the anomaly score used for detection. Intuitively, if a system cannot reliably distinguish between machines, it is unlikely to detect subtle deviations from machine-specific normal sounds.

The observed degradation can also be interpreted from a ranking perspective. When machine identity is unknown, each test recording is compared against multiple machine-specific notions of normality, and errors occur when an incorrect machine yields a lower anomaly score than the true one. This reflects overlap between machine-specific score distributions. Rank-based metrics such as AUC and pAUC capture this behavior directly, explaining the strong empirical relationship between anomaly detection performance degradation and implicit machine identification accuracy. These results suggest that robustness to missing machine identity should be considered an explicit evaluation dimension in future \ac{asd} benchmarks.

Finally, the proposed evaluation setting has several limitations. We assume single-channel recordings containing sounds from a single machine and a fixed set of machines. While consistent with current DCASE benchmarks, these assumptions limit applicability to more complex deployments. In practice, recordings may contain multiple simultaneously operating machines and overlapping normal and anomalous events, requiring systems to identify all active machines and produce machine-specific anomaly scores. Extending the evaluation to such scenarios is an important direction for future work.

\section{Conclusion}
In this work, we investigated how much machine identity matters for anomalous sound detection at test time by introducing a minimal modification to the standard DCASE evaluation protocol that removes access to machine identity during inference. We showed that this assumption has a substantial impact on reported performance and that relaxing it reveals clear differences in robustness hidden by standard machine-wise evaluation. Importantly, machine identity does not matter uniformly across methods: Approaches that can implicitly recover the correct machine-specific notion of normality are largely unaffected, whereas methods relying on explicit machine-specific models degrade due to ambiguity in score selection and overlapping score distributions. We further demonstrated a strong relationship between anomaly detection performance degradation and implicit machine identification accuracy, providing a simple explanation of when and why machine identity matters and motivating its consideration as an explicit evaluation dimension in future \ac{asd} benchmarks.

\section{Generative AI disclosure}
Generative AI tools were used for language editing and polishing of the manuscript.
All scientific content, interpretations, and conclusions are the responsibility of the authors.

\bibliographystyle{IEEEtran}
\bibliography{refs}

\end{document}